\documentclass[epjST]{svjour}
\usepackage{graphicx}
\usepackage{bm}
\newcommand{\boldsigma}{\ensuremath{\bm{\sigma}}}

\newcommand{\boldrho}{\ensuremath{\bm{\rho}}}
\newcommand{\boldeps}{\ensuremath{\bm{\epsilon}}}

\newcommand{\mtx}[4]{\left(\begin{array}{cc}{#1}&{#2}\\{#3}&{#4}\end{array}\right)}

\newcommand{\tr}[0]{\mathrm{tr}}

\newcommand{\chem}[1]{\ensuremath{\mathrm{#1}}}
\newcommand{\un}[1]{\ensuremath{\,\mathrm{#1}}}
\begin{document}
\title{X-ray polarization: General formalism and polarization analysis}
\titlerunning{X-ray polarization I}
\author{C. Detlefs\inst{1}\and M. Sanchez del Rio\inst{1} \and C. Mazzoli\inst{2}}
\authorrunning{Detlefs and Mazzoli}
\institute{European Synchrotron Radiation Facility, B.P. 220, 6, rue Jules Horowitz, F--38043 Grenoble Cedex, France
	\and
	Politecnico di Milano, Dipartimento di Fisica, Piazza Leonardo da Vinci 32, 20133 Milano, Italy
}
\abstract{
	The polarization of x-rays plays an outstanding role in experimental 
	techniques such as non-resonant magnetic x-ray scattering and resonant 
	x-ray scattering of magnetic and multipolar order. Different instrumental
	methods applied to synchrotron light can transform its natural polarization 
	into an arbitrary polarization state. Several synchrotron applications, in 
	particular in the field of magnetic and resonant scattering rely on 
	the improvement in the signal/noise ratio or the deeper insight into the ordered 
	state and the scattering process made possible through these polarization techniques. 
	Here, we present the mathematical framework for the description of fully and partially 
	polarized x-rays, with some applications such as linear x-ray polarization analysis for 
	the determination of the scattered beam's polarization, and the $\mathrm{Ge}$ K-edge 
	resonant scattering.
} %end of abstract
\maketitle
%
%%%%%%%%%%%%%%%%%%%%%%%%%%%%%%%%%%%%%%%%%%%%%%%%%%%%%%%%%%%%%%%%%%%%%%%%%%%%%%%%%%%%

%\section{Introduction}
%\section{Scope}

\section{Mathematical description of polarized light}

Previous reviews of x-ray polarization phenomena are mostly focused on aspects related to
crystal dynamical diffraction \cite{Hart78,Belyakov89}. Several publications 
discuss the polarization dependent scattering amplitudes for non-resonant magnetic
\cite{deBergevin72,Blume88} and magnetic and higher order multipolar resonant 
scattering \cite{Hill96b}, without introducing the formalism.
Theoretical work on the polarization dependence of the resonant cross section mostly 
uses a tensor notation that is beyond the scope of this paper 
\cite{Hannon88,Carra93,Lovesey96c,DiMatteo03}. We here present a description of x-ray 
polarization based on classical electrodynamics with a view towards applications in 
magnetic scattering research with synchrotron radiation.

X-rays are transverse electro-magnetic waves \cite{Barkla06}, just like
visible light, so that the description of polarized optics in the visible regime
may be applied \cite{Born99,Goldstein11}. The polarization state of an isolated
wave with wave vector $\vec{k}$ and photon energy $\hbar\omega$,
\begin{equation}
        \vec{E}(t, \vec{r})
        = \Re\left[
                \left(V_1 \hat{\boldeps}_1 
                + V_2 \hat{\boldeps}_2\right)
        \cdot 
        e^{-i(\omega t - \vec{k}\cdot\vec{r})}
        \right]
        \label{el_field}
\end{equation}
is completely defined by the components of the Jones vector,
$\vec{V} = \left(\begin{array}{c}V_1\\V_2\end{array}\right)$. 
The components $V_1$ and $V_2$ may be complex, e.g., in the case of circular
polarization. The coordinate system $\hat{\boldeps}_{1,2,3}$ is
chosen such that $\hat{\vec{k}}=\hat{\boldeps}_3$,
$\hat{\boldeps}_1 \times \hat{\boldeps}_2 =
\hat{\boldeps}_3$ and $\hat{\boldeps}_2 \times
\hat{\boldeps}_3 = \hat{\boldeps}_1$.

For the purpose of this paper we define the intensity of the
beam as
$
  I = \left|V_1\right|^2
  + \left|V_2\right|^2,
$
dropping the usual proportionality factors.

\subsection{Polarization Ellipse}

\begin{figure}\sidecaption
	\mbox{%
		\includegraphics[width=0.5\columnwidth]{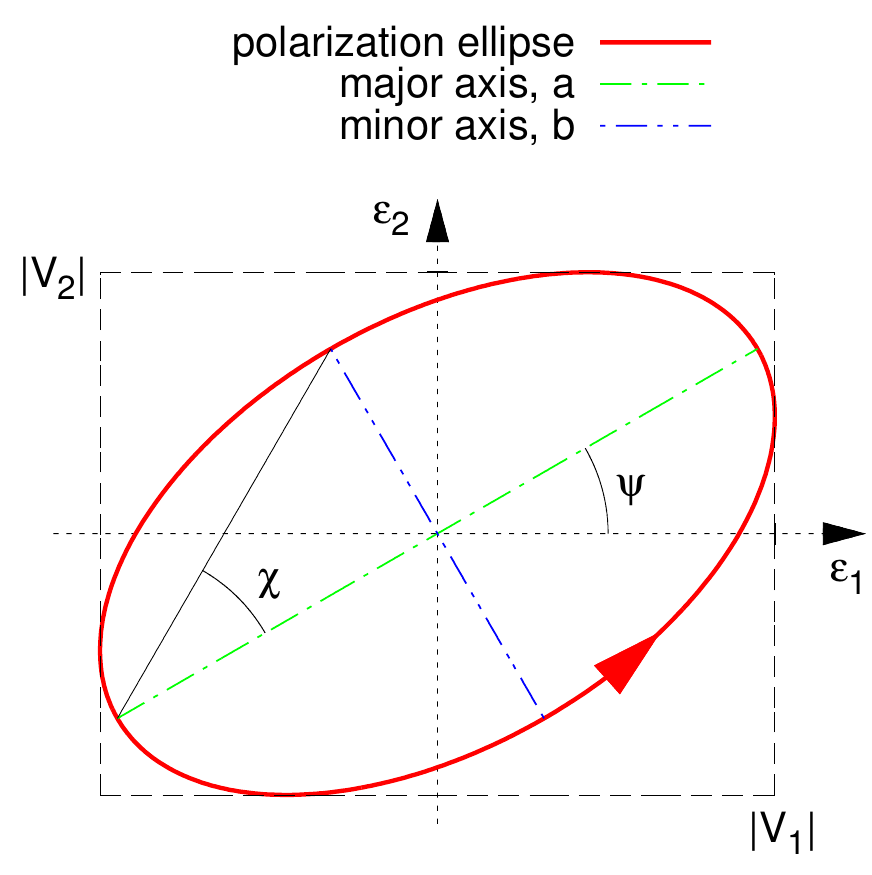}
	}
	\caption{\label{fig_poln_ellipse}The polarization ellipse of a general 
		(elliptically polarized) beam. In this example $\psi=\chi=30^\circ$, 
		yielding $P_1={1}/{4}$, $P_2={\sqrt{3}}/{4}$ and $P_3={\sqrt{3}}/{2}$. 
		For left-handed polarization ($P_3>0$), the electric field vector rotates 
		counterclockwise when looking into the source.}
\end{figure}

Following \cite{Goldstein11}, we write eq.~\ref{el_field} as
\begin{equation}
	\frac{E_i(\tau)}{\left| V_i \right|} = \cos(\tau) \cos(\phi_i) + \sin(\tau) \sin(\phi_i)
\end{equation}
where $i=1,2$. $\tau = \omega t - \vec{k}\cdot\vec{r}$ is called the propagator, 
$\left| V_i \right|$ are the maximum amplitudes, and $\phi_i$ the phases 
of the two polarizations, such that $V_{i} = \left| V_i \right| \exp(i \phi_i)$. 
Eliminating the explicit dependence on the propagator we obtain \cite{Goldstein11}:
\begin{equation}
	\frac{E^2_1(\tau)}{\left| V_1\right|^2} + 
	\frac{E^2_2(\tau)}{\left| V_2\right|^2} -
	2 \frac{E_1(\tau) E_2(\tau)}{\left| V_1 \right| \left| V_2 \right|} \cos(\Delta\phi) 
	=
	\sin^2(\Delta\phi),
	\label{eq_ellipse}
\end{equation}
where $\Delta\phi = \phi_2 - \phi_1$ is the relative phase between the two components of 
the polarized beam. 
We see that the projection of all instantaneous electric field vectors onto a plane 
perpendicular to the direction of propagation $\hat{\vec{k}}$ fall onto an ellipse, 
the \emph{polarization ellipse} \cite{Goldstein11} as shown in Fig.~\ref{fig_poln_ellipse}.

The major ($2a$) and minor ($2b$) axes of the ellipse are given by the equations \cite{Goldstein11} 
\begin{eqnarray}
	a^2 + b^2 & = & \left| V_1\right|^2 + \left| V_2\right|^2 \\
	a b & = & \left| V_1 \right| \left| V_2 \right| \left| \sin(\Delta\phi) \right|,
\end{eqnarray} 
and the angle $\psi$ between the major axis and coordinate axis $\hat{\boldeps}_1$ 
is given by
\begin{equation}
	\tan(2\psi) 
		= 
		2 \frac{
			\left| V_1 \right| \left| V_2 \right|
		}{
			\left| V_1 \right|^2 - \left| V_2 \right|^2
		} \cos(\Delta\phi),
	\label{eq_psi}
\end{equation}
with $0 \leq \psi \leq \pi$. Finally, the angle of ellipticity $\chi$ is defined by 
\begin{equation}
	\sin(2\chi) 
	= 
	2 \frac{
		\left| V_1 \right| \left| V_2 \right|
	}{
		\left| V_1 \right|^2 + \left| V_2 \right|^2
	} \sin(\Delta\phi),
	\label{eq_chi}
\end{equation}
with $- \frac{\pi}{4} \leq \chi \leq \frac{\pi}{4}$ so that 
$\tan(\left|\chi\right|)=b/a$ (see Fig.~\ref{fig_poln_ellipse}). 
%The angles $2\chi$ and $2\psi$ 
%can be visualized as polar coordinates of a point on the surface of the Poincar\'e 
%sphere \cite{Born99,Goldstein11}.
%, see below and Fig.~\ref{fig.poincare_sphere}.

\subsection{Degenerate cases of the polarization ellipse and special polarizations}

A beam with vanishing minor axis, $b=0$, is called \emph{linearly polarized}. In this case 
the two polarizations oscillate in phase, $\Delta\phi=0$, or in antiphase, 
$\Delta\phi=\pi$. The angle of ellipticity is $\chi=0$, and the corresponding 
polarization ellipse collapses to a line.

A beam with $\sin(2\chi) = \pm 1$, has $\sin(\Delta\phi) = \pm 1$, 
i.e.~$\Delta\phi = \pm \pi/2$, and $\left|V_{1}\right| = \left|V_{2}\right|$ 
and therefore $a=b$. In this case the angle $\psi$ is undefined.
The corresponding polarization ellipse is a circle, and the beam has left- ($\Delta\phi=\pi/2$) 
or right-handed ($\Delta\phi=-\pi/2$) \emph{circular polarization}, with electric field vectors 
rotating counterclockwise and clockwise, respectively, when looking into the source\footnote{In 
the literature, the nomenclature 
for left- or right-handedness and for positive or negative helicity is not always consistent. We 
derive the sign of the helicity from eq.~\ref{trace.polarization}.
An example for a left-handed circular wave with $P_3=+1$ 
is  $V_1={1}/{\sqrt{2}}$ and $V_2={i}/{\sqrt{2}}$. The phase and amplitude of this vector may be 
changed to obtain other waves with left-handed circular polarization}.

\subsection{Poincar\'e-Stokes parameters}

The Poincar\'e-Stokes parameters $P_1$, $P_2$, and $P_3$ completely
describe the state of polarization of a beam. They are defined as
\begin{eqnarray}
  P_1
  & = & 
  \frac{
    \left|V_1\right|^2 - \left|V_2\right|^2
    }{
    \left|V_1\right|^2 + \left|V_2\right|^2
    }
\\
  P_2 
  & = & 
  \frac{
    \left|V_1+V_2\right|^2 - \left|V_1-V_2\right|^2
    }{
    2 \left(\left|V_1\right|^2 + \left|V_2\right|^2\right)
    }
  \\
  P_3
  & = & 
  \frac{
    \left|V_1- i V_2\right|^2 - \left|V_1+ i V_2\right|^2
    }{
    2 \left(\left|V_1\right|^2 + \left|V_2\right|^2\right)
    }.
\end{eqnarray}

$P_1$ and $P_2$ describe the state of linear polarization, 
and $P_3$ (sometimes called $P_{\mathrm{c}}$) the degree of circular polarization
\cite{Born99,Goldstein11,Stokes52,deBergevin81a}, 
with $P_3=+1$ for left-handed circular polarization. In addition, we 
define the degree of linear polarization as $P_{\mathrm{lin}}=\sqrt{P_1^2+P_2^2}$
%Note that we follow the notation of \cite{Born99},
%\cite{deBergevin81a} and \cite{Lipps54} rather than that of
%\cite{Blume88} and \cite{Lovesey01}. However, this affects neither the
%notation of the density matrix (see below) nor that of the Jones matrices (see
%section~\ref{sec.jones}).  
$\mathbf{P} = (P_1, P_2, P_3)$ is called
the Poincar\'e-Stokes polarization vector, although it does not have
the transformation properties of a vector ($P_{1,2}$ have even parity,
while $P_3$ has odd parity.  $P_{1,2,3}$ are even under time reversal
\cite{Lovesey01}). It is related to the Stokes vector, $\vec{S}$, more commonly
used in optics via $S_0=I$, $S_{1,2,3}=I P_{1,2,3}$
\cite{Born99,Goldstein11}. 

An isolated photon is 100\% polarized with
$P = \sqrt{P_1^2+P_2^2+P_3^2} = 1$. The 
Poincar\'e-Stokes parameters are related to the polarization ellipse, 
eqs.~\ref{eq_psi} and~\ref{eq_chi}, as follows 
(see Fig.~\ref{fig_poln_ellipse}):
% and~\ref{fig.poincare_sphere}):
\begin{eqnarray}
	P_1 & = & \cos(2\psi) \cos(2\chi) \label{stokes_ellipse_1} \\
	P_2 & = & \sin(2\psi) \cos(2\chi) \label{stokes_ellipse_2} \\
	P_3 & = & \sin(2\chi) \label{stokes_ellipse_3}.
\end{eqnarray}

\subsection{Coherency matrix}
\label{section.coherency}

An X-ray beam that is composed of an ensemble of independent waves
may be partially polarized, i.e.~$P < 1$.  Such a beam cannot be
described by a simple Jones vector.  Instead, its
polarization state can be described by a density matrix
\cite{Blume88,deBergevin81a,Lipps54,Fano57,Lipps54b}, in analogy to
the coherency matrix used in classical optics \cite{Born99,Goldstein11}.

\begin{equation}
  \boldrho = \langle \vec{V}\vec{V}^\dagger \rangle = \mtx{%
    \langle V_1 V_1^\dagger \rangle}{%
    \langle V_1 V_2^\dagger \rangle}{%
    \langle V_2 V_1^\dagger \rangle}{%
    \langle V_2 V_2^\dagger \rangle}
=
\frac{I}{2} \left(\mathbf{1}+ \boldsigma \cdot \mathbf{P}\right),
\label{eq_coherency}
\end{equation}
where the average, $\langle \ldots \rangle$, is taken over the ensemble of waves
constituting the beam \cite{Born99}. Depending on the 
experiment the average may require integrals over linear (beam size) and angular space 
(divergence), energy (bandwidth) and time (fluctuations), 
as the polarization may depend on each one of these parameters.
 $\boldsigma = (\boldsigma_1,
\boldsigma_2, \boldsigma_3)$ represents the Pauli matrices,
$\boldsigma_1 = \mtx{1}{0}{0}{-1}$, $\boldsigma_2=\mtx{0}{1}{1}{0}$,
and $\boldsigma_3=\mtx{0}{-i}{i}{0}$. 

%\begin{figure}
%  \centerline{%
%    \includegraphics[width=0.7\columnwidth]{fig_poincare_sphere}
%  }
%  \caption{\label{fig.poincare_sphere}Graphical representation of the
%  	Poincar\'e-Stokes vector in the Poincar\'e sphere. For a fully polarized beam 
%  	($P=1$, i.e.~\vec{P} falls onto the surface of the sphere) the angle between 
%  	the projection of $\vec{P}$ onto the equatorial plane and the $P_1$ axis 
%  	is twice the angle between the major axis of the polarization ellipse and 
%  	the $\hat{\boldeps}_1$ axis, $\psi$. The angle between $\vec{P}$ and the 
%  	equatorial plane is twice the angle of ellipticity, $\chi$. See
%  	eqs.~\ref{stokes_ellipse_1}--\ref{stokes_ellipse_3} and Fig.~\ref{fig_poln_ellipse}.}
%\end{figure}

The intensity and Poincar\'e-Stokes parameters for the X-ray beam above 
are easily extracted from the coherency matrix,
\begin{eqnarray}
  I & = & \tr(\boldrho) \label{trace.intensity}\\
  P_{i} & = & \frac{1}{I}\tr(\boldsigma_{i}\cdot \boldrho),\ i=1,2,3
  \label{trace.polarization}.
\end{eqnarray}

%\subsection{Poincar\'e sphere}
%
%Graphically, the state of polarization can be represented as a point
%inside the Poincar\'e sphere \cite{Born99,Goldstein11} (see
%Fig.~\ref{fig.poincare_sphere}). Points on the surface of the sphere
%represent fully polarized beams, $P=1$. In particular, the poles
%represent complete left- ($P_3=+1$, $P_{1,2}=0$) or right-handed ($P_3=-1$) circular 
%polarization, whereas points on the equator represent linearly polarized 
%beams ($P_{\mathrm{lin}}=\sqrt{P_1^2+P_2^2}=1$, $P_3=0$). Partially polarized 
%beams with $P<1$ fall within the sphere, with completely unpolarized beams 
%($P=0$) being represented by the center of the sphere.

\subsection{Choice of coordinate system}

\begin{figure}\sidecaption
  \mbox{%
    \includegraphics[width=0.5\columnwidth]{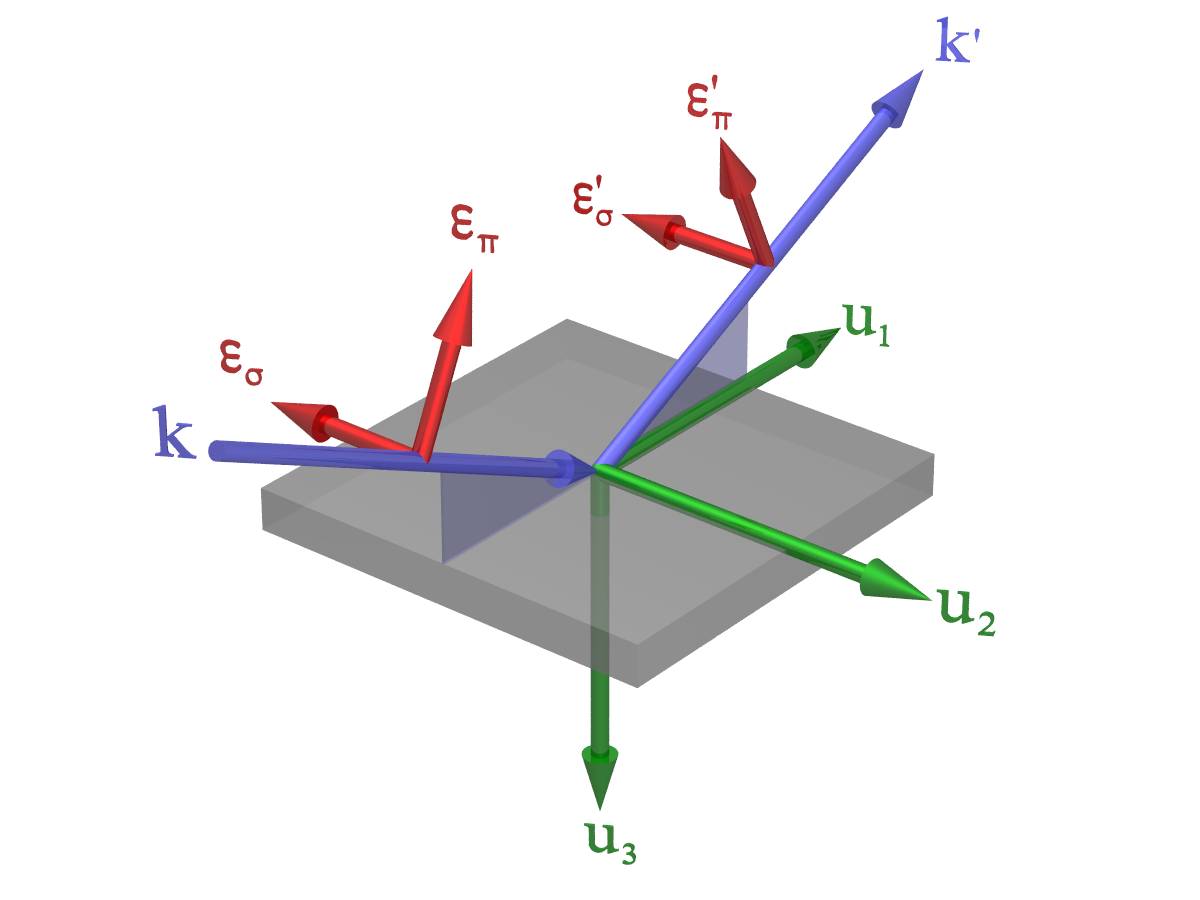}
    }
  \caption{\label{fig.coord}Coordinate system and polarization
    vectors relative to the incident ($\vec{k}$) and scattered 
    ($\vec{k}^\prime$) beams.}
\end{figure}

For an isolated wave, the choice of the coordinate axes
$\hat{\boldeps}_{1,2}$ is arbitrary -- $\hat{\boldeps}_3$
is given by the beam direction. When describing a scattering process 
these axes are usually chosen to be perpendicular ($\sigma$)
and parallel ($\pi$) to the scattering plane.
Following \cite{Blume88} and \cite{Hill96b}, we define the coordinate 
system $\hat{\vec{u}}_{1,2,3}$ through the
incident and scattered beam directions, $\hat{\vec{k}}$ and $\hat{\vec{k}}^\prime$
(see Fig.~\ref{fig.coord}).
\begin{eqnarray}
        \hat{\vec{u}}_1 
& = &   \frac{1}{2\cos(\theta)}
        \left(\hat{\vec{k}}+\hat{\vec{k}}^\prime\right)
\label{eq:coord_u1} \\
        \hat{\vec{u}}_2
& = &   \frac{1}{\sin(2\theta)}
        \left(\hat{\vec{k}}\times\hat{\vec{k}}^\prime\right)
\label{eq:coord_u2} \\
        \hat{\vec{u}}_3
& = &   \frac{1}{2\sin(\theta)}
        \left(\hat{\vec{k}}-\hat{\vec{k}}^\prime\right),
\label{eq:coord_u3} 
\end{eqnarray}
where $2\theta$ is the scattering angle 
($\cos(2\theta) = \hat{\vec{k}} \cdot \hat{\vec{k}}^\prime$). 
We choose the polarization vectors as
\begin{eqnarray}
        \hat{\boldeps}_{\sigma} = \hat{\boldeps}_1 
& = &   -\hat{\vec{u}}_2
\label{eq:coord_eps1} \\
        \hat{\boldeps}_{\pi} = \hat{\boldeps}_2 
& = &   \sin(\theta) \hat{\vec{u}}_1 - \cos(\theta) \hat{\vec{u}}_3
\label{eq:coord_eps2}\\
        \hat{\boldeps}_{\sigma}^\prime = \hat{\boldeps}_1^\prime 
& = &   -\hat{\vec{u}}_2
\label{eq:coord_eps1prime}\\
        \hat{\boldeps}_{\pi}^\prime = \hat{\boldeps}_2^\prime
& = &   -\sin(\theta) \hat{\vec{u}}_1 - \cos(\theta) \hat{\vec{u}}_3.
\label{eq:coord_eps2prime}
\end{eqnarray}
$\hat{\boldeps}_1$ and $\hat{\boldeps}^\prime_1$ are
perpendicular to the scattering plane ($\sigma$ polarization), while
$\hat{\boldeps}_2$ and $\hat{\boldeps}_2^\prime$ lie
within the scattering plane ($\pi$ polarization). In the following the
indices $1$ and $2$ may be replaced by $\sigma$ and $\pi$,
respectively, when the discussion is restricted to a single scattering
process with a well defined scattering plane.

\begin{figure}
	\centerline{%
		\includegraphics[width=0.70\columnwidth]{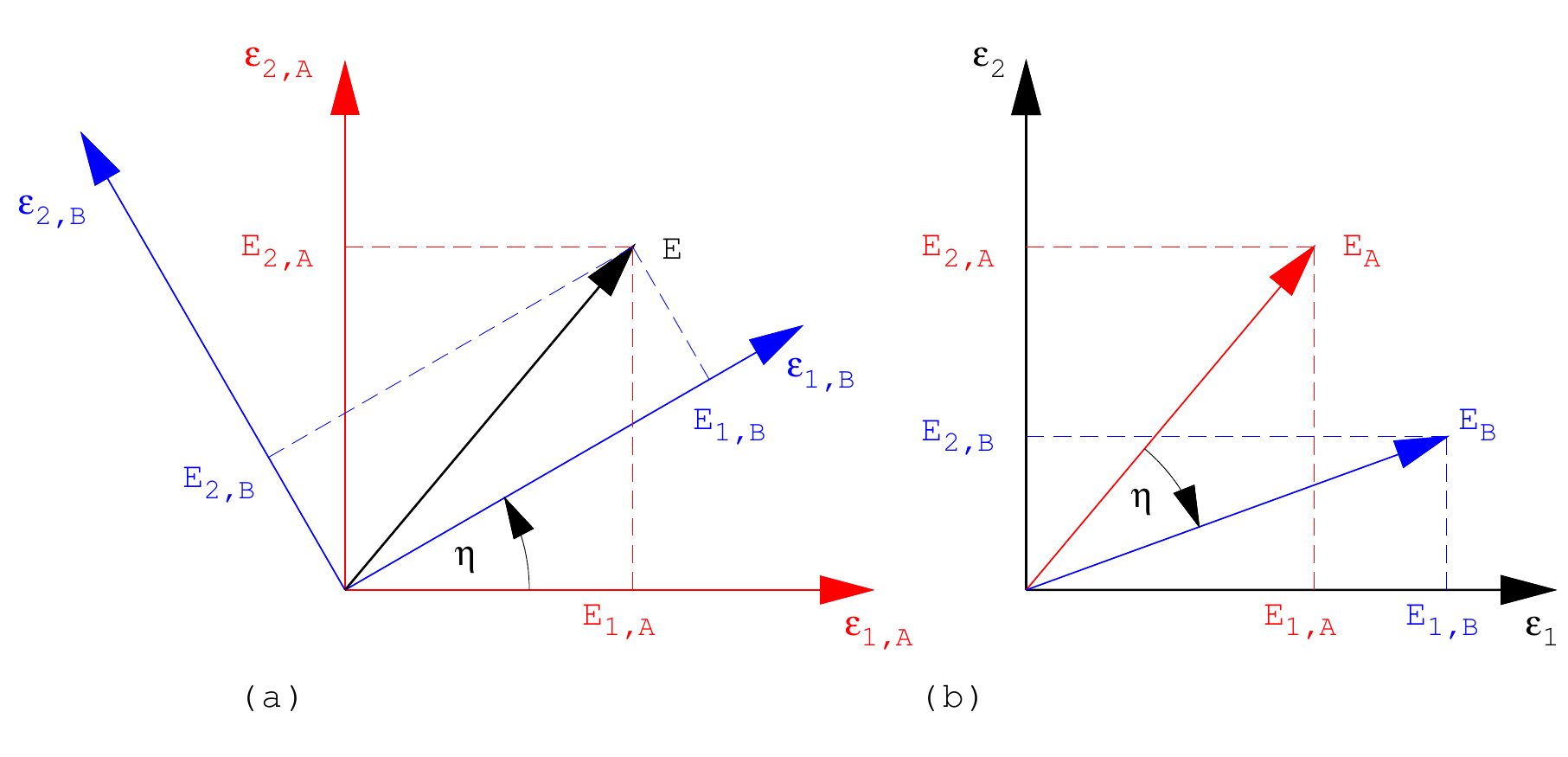}
	}
	\caption{
		\label{fig_coordtransform}A change of the coordinate system from 
		$A$ to $B$ is effected by an orthogonal rotation matrix. (a) Transformation of the coordinate 
		system relative to a fixed laboratory reference. (b) Transformation of the vector components 
		in these coordinate systems.
	}
\end{figure}

In magnetic scattering one frequently encounters subsequent
diffraction processes with different diffraction planes, such as
vertical diffraction in the monochromator, horizontal diffraction from
the sample, and vertical diffraction in a polarization analyzer.
This change of reference system is accomplished through an orthogonal
transformation \cite{Born99},
$R(\eta)=\mtx{\cos(\eta)}{\sin(\eta)}{-\sin(\eta)}{\cos(\eta)}$. Here 
$\eta$ is the rotation angle between the two reference systems
$A$ and $B$, such $\vec{E}_B(\tau) = R(\eta) \cdot \vec{E}_A(\tau)$, 
therefore $\vec{V}_B = R(\eta)\cdot \vec{V}_A$ and 
$(\hat{\boldeps}_{1,B}, \hat{\boldeps}_{2,B}) = 
(\hat{\boldeps}_{1,A}, \hat{\boldeps}_{2,A})\cdot R^{-1}(\eta)$,
see Fig.~\ref{fig_coordtransform}. 
The density matrix transforms as 
$\boldrho_B = R(\eta) \cdot \boldrho_A \cdot R^{-1}(\eta)$, 
yielding the Poincar\'e-Stokes parameters
\begin{eqnarray}
        P_{1,B} 
& = &   \cos(2\eta) P_{1,A} + \sin(2\eta) P_{2,A} 
\label{eq.rot_p1} \\
        P_{2,B}
& = &   -\sin(2\eta) P_{1,A} + \cos(2\eta) P_{2,A} 
\label{eq.rot_p2} 
\end{eqnarray}

The degree of linear polarization, $P_{\mathrm{lin}}$, 
is invariant under this transformation. Also
invariant are the intensity, $I$, the circular polarization, $P_3$ and
the total degree of polarization, $P$.

$P_{1,B}$ is maximized and $P_{2,B}=0$ when the major axis 
of the polarization ellipse is rotated onto the $\hat{\boldeps}_1$ axis, 
i.e.~for $\sin(2\eta)={P_{2,A}}/{P_{\mathrm{lin}}}$ and 
$\cos(2\eta)={P_{1,A}}/{P_{\mathrm{lin}}}$, see~eq.~\ref{eq_psi} 
and Fig.~\ref{fig_poln_ellipse}.

\subsection{Jones Matrices}
\label{sec.jones}

Jones matrices are used to describe the effect of an optical element 
on the beam. This formalism can be used to describe 
diffracting elements, filters and dichroic samples.

Let the operator $\mathbf{M}$ describe the (forward) scattering of an
X-ray optical element, such as the phase plate, a sample, a
polarization analyzer, or any other optical element in the path of the
beam. The effect of $\mathbf{M}$ on the beam is completely described
by the matrix elements of the basis vectors of polarization. They may
be conveniently written in form of a $2\times2$ matrix, the Jones
matrix \cite{Jones41,Hurwitz41,Jones41b},
\begin{equation}
  \mathbf{M} = \mtx{%
    \left<\hat{\boldeps}_\sigma^\prime\right| M
    \left|\hat{\boldeps}_\sigma\right> 
    }{%
    \left<\hat{\boldeps}_\sigma^\prime\right| M
    \left|\hat{\boldeps}_\pi\right> 
    }{%
    \left<\hat{\boldeps}_\pi^\prime\right| M
    \left|\hat{\boldeps}_\sigma\right> 
    }{%
    \left<\hat{\boldeps}_\pi^\prime\right| M
    \left|\hat{\boldeps}_\pi\right> 
    },
\end{equation}
such that $\vec{V}^\prime = \mathbf{M} \cdot \vec{V}$. Consider the 
aforementioned coordinate transformation by an orthogonal matrix $R$:
\begin{equation}
	  \underbrace{R \cdot \vec{V}_A^\prime}_{\vec{V}_B^\prime} 
  = 
	 	\underbrace{R \cdot \mathbf{M}_A \cdot R^{-1}}_{\mathbf{M}_B} 
	 	\cdot \underbrace{R \cdot \vec{V}_A}_{\vec{V}_B},
\end{equation}

The same Jones matrix may be used with the coherency matrix formalism 
(section \ref{section.coherency}), 
so that the effect of a scattering process can also be studied for 
partially polarized beams.
The density matrix of the scattered beam is then given by
\begin{equation}
        \boldrho^\prime 
 =    \mathbf{M} \cdot \boldrho \cdot \mathbf{M}^\dagger.
\label{eq.defscatteredbeam}
\end{equation}
The intensity, $I^\prime$, and Poincar\'e-Stokes parameters, $\mathbf{P}^\prime$,
of the scattered beam are easily obtained using 
equations~\ref{trace.intensity} and~\ref{trace.polarization}.

Subsequent diffraction off several optical elements, $\mathbf{M}_1,
\mathbf{M}_2, \ldots$ (e.g., a monochromator, a phase plate, a sample,
and a polarization analyzer) are represented by the product of the
Jones matrices of the individual scatterers,
\begin{eqnarray}
  \rho^\prime 
  & = & 
  \left(\ldots \mathbf{M}_2 \cdot \mathbf{M}_1\right)
  \cdot \rho \cdot
  \left(\ldots \mathbf{M}_2 \cdot \mathbf{M}_1\right)^\dagger
%  \\
%  & = &
	=
  \ldots \left[\mathbf{M}_2 \cdot \left(\mathbf{M}_1
  \cdot \rho \cdot
  \mathbf{M}_1^\dagger\right) \cdot \mathbf{M}_2^\dagger\right] \cdot \ldots.
\end{eqnarray}

The possible polarization sensitivity of a detector can be described
by an additional Jones matrix acting as a polarization
filter. Standard X-ray detectors, however, are not sensitive to the
polarization of the detected radiation. Instead, polarization
analyzers based on Bragg diffraction are used (see section
\ref{examples_pa}).

The Jones matrices for non-resonant magnetic scattering are given in
\cite{Blume88}, and those for resonant magnetic and
ATS scattering have been tabulated in \cite{Hill96b}.

\subsection{M{\"u}ller matrices}

The net effect of an ensemble of scatterers on the intensity and 
polarization can be elegantly described using the M{\"u}ller calculus
\cite{Born99,Goldstein11}.
\begin{equation}
		\vec{S}^\prime 
  =
		{\cal M} \cdot \vec{S}, 
\end{equation}
where ${\cal M}$ is the M{\"u}ller 
matrix, and $S_{i} = \tr(\boldsigma_i \cdot \boldrho)$ for $i=0,1,2,3$ 
is the Stokes vector with $\boldsigma_0 = \mtx{1}{0}{0}{1}$ (see above).

The M{\"u}ller  matrix can be derived from the corresponding Jones matrix, 
$\mathbf{M}$ \cite{Kim87}:
\begin{equation}
		{\cal M}_{j,i} 
	=  
		\frac{1}{2} \tr(\boldsigma_i \cdot \mathbf{M} \cdot \boldsigma_j \cdot \mathbf{M}^\dagger).
		\label{eq_jones_to_mueller}
\end{equation}
The Jones matrix, however, contains the phase of the scattering process,
therefore it cannot be determined from the M{\"u}ller matrix which does
not contain the phase information.
The advantage of the M{\"u}ller calculus is that it can also describe 
depolarizing elements. This is not possible with a single Jones matrix.

In the case of a large beam described by $\boldrho$ that diffracts of 
an ensemble of independent scatterers, e.g.~different magnetic domains or 
grains of a mosaic crystal, the density matrix has to be calculated 
for each scatterer.
Let $n$ enumerate the scatterers with probability $p(n)$, then 
$\boldrho^\prime_n  =  \mathbf{M}_n \cdot \boldrho \cdot \mathbf{M}_n^\dagger$. 
The composite scattered beam is then described by the sum of these
density matrices, $\boldrho^\prime = \sum_n p(n) \boldrho^\prime_n$.
Note that one does \emph{not} obtain the same result by averaging over the
individual Jones matrices!

The net effect of such an ensemble of scatters can be described by a 
single  M{\"u}ller matrix,
${\cal M}_{ij} = \frac{1}{2} \sum_n p(n) \tr(\boldsigma_i \mathbf{M}_n \boldsigma_j \mathbf{M}_n^\dagger)$, 
which is the average of the  M{\"u}ller matrices describing the individual 
scattering processes, ${\cal M} = \sum_n p(n) {\cal M}_n$ \cite{Kim87}.

%%%%%%%%%%%%%%%%%%%%%%%%%%%%%%%%%%%%%%%%%%%%%%%%%%%%%%%%%%%%%%%%%%%%%%%%%%%

\section{Thomson scattering and linear polarization analyzer}
\label{sec.examples.thomson}

Thomson scattering is the elastic scattering by a free charged particle. 
It arises from the isotropic (scalar) polarizability
of the scatterer, $f = f(\vec{Q})\ \hat{\boldeps}^{\prime\dagger}
\cdot \hat{\boldeps}$, where $f(\vec{Q})$ is the form factor, 
i.e.~the Fourier transform of the particle's charge distribution.
Thomson scattering can be observed in many classes of samples, such as 
gases, liquids, amorphous solids, single crystals or polycrystalline powders.
In the context of resonant or anomalous scattering, 
Thomson scattering is referred to as ``normal'' or ``charge'' scattering. 
The corresponding  Jones matrix is given by 
(using the coordinate system shown in Fig.~\ref{fig.coord})
\begin{equation}
        \mathbf{M}_{\mathrm{Th}} = \alpha
        \mtx{1}{0}{0}{\cos(2\theta)},
        \label{jones-thomson}
\end{equation}
where $2\theta$ is the scattering angle. 
The proportionality constant $\alpha$ contains all terms that 
are independent of the polarization, e.g.~the form factor $f(\vec{Q})$ 
at the chosen scattering vector $\vec{Q} = \vec{k}^\prime - \vec{k}$ 
in disordered systems, the structure factor $F(\vec{Q})$ in crystals, 
 and geometrical terms \cite{Warren69}. Note that eq.~\ref{jones-thomson} 
holds only In the kinematic approximation, i.e.~in the 
absence of multiple scattering and anomalous effects. 

When an X-ray beam undergoes Thomson scattering, its Stokes
parameters are modified as follows \cite{Ramaseshan53}:
\begin{eqnarray}
        \boldrho^\prime 
& = &   \mathbf{M}_{\mathrm{Th}} \boldrho \mathbf{M}_{\mathrm{Th}}^\dagger
%\\
%& = &  
=
				\alpha \mtx{%
                1+P_1}{%
                \left(P_2-iP_3\right)x}{%
                \left(P_2+iP_3\right)x}{%
                \left(1-P_1\right)x^2}
        \label{eq.th_rho}
\\
        I^\prime
& = &  
				\frac{I}{2} \left|\alpha\right|^2 \left[%
                1 + x^2 + P_1 (1-x^2)%
        \right]
        \label{eq.th_int}
\\
        P_1^\prime
& = &   \frac{%
                1 - x^2 + P_1 (1+x^2)
        }{%
                1 + x^2 + P_1 (1-x^2)
        }
        \label{eq.th_p1}
\\
        P_2^\prime
& = &
        \frac{%
                2 P_2 x%
        }{%
                1 + x^2 + P_1 (1-x^2)
        }
        \label{eq.th_p2}
\\
        P_3^\prime
& = &
        \frac{%
                2 P_3 x%
        }{%
                1 + x^2 + P_1 (1-x^2)
        },
        \label{eq.th_p3}
\end{eqnarray}
where $x=\cos(2\theta)$.

%\subsection{$\cos(2\theta) \approx 0$: Linear polarization analyzer (PA)}
\label{examples_pa}

For $\cos(2\theta)=0$, $\mathbf{M}=\alpha \mtx{1}{0}{0}{0}$ (eq.~\ref{jones-thomson}), 
i.e.~only the $\sigma$ component survives. The Stokes parameters
(eqs.~\ref{eq.th_int}--\ref{eq.th_p3}) of the scattered beam 
reduce to
\begin{eqnarray}
        I^\prime 
& = &   \frac{I}{2} \left|\alpha\right|^2 \left(1+P_1\right)
\\
        P_1^\prime
& = &   1 \\
        P_{2,3}^\prime
& = &   0,
\end{eqnarray}
except for the case $P_1=-1$, when the intensity of the scattered beam
vanishes, $I^\prime=0$.

\begin{figure}
  \centerline{%
    \includegraphics[width=0.85\columnwidth]{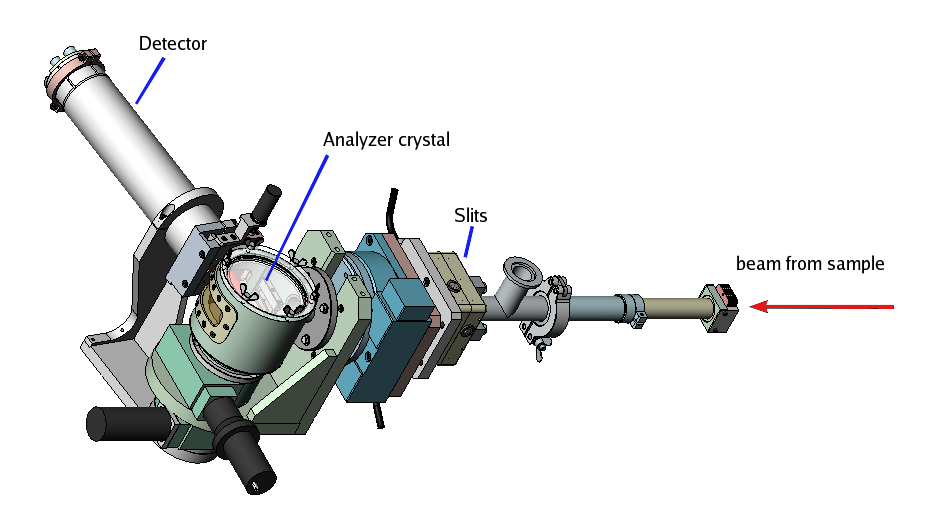}
    }
  \caption[]{\label{fig_pa}Experimental configuration for linear 
    polarization analysis of X-rays at the ESRF Beamline ID20 \cite{Paolasini07}. 
    The beam scattered 
    off the sample enters an evacuated flight tube (right). In-vacuum slits 
    are positioned upstream of the analyzer crystal. The analyzer crystal 
    and detector are mounted on a common goniometer rotating about the beam 
    ($\eta$-axis). The Bragg ($\theta_{\mathrm{PA}}$) and scattering angles 
    ($2\theta_{\mathrm{PA}}$) can be adjusted individually. }
\end{figure}

A Thomson scatterer with $2\theta=90^\circ$, and that can be rotated
by an angle $\eta$ about the beam axis, may thus be used as a
linear polarization analyzer (PA) to determine the linear polarization
parameters, $P_1$ and $P_2$ (see Fig.~\ref{fig_pa}). 
Note that the degree of circular polarization, $P_3$, cannot
be measured in this way. It is, however, possible to establish an 
upper bound, as $P_3^2 \leq 1 - P_1^2 - P_2^2$.

Scatterers may be
incoherent, e.g., an amorphous Kapton foil, or Bragg scatterers, in
which case the condition $\cos(2\theta)=0$ is generally fulfilled only
approximatively, as a perfect match of the PA crystal's $d$-spacing 
to the desired photon 
energy cannot always be obtained (see table~\ref{tab.pacrystals}). 
In practice there is always some crosstalk between
the polarization channels, $0 < x \ll 1$, generally referred to as
``spillover''.

Typically only the intensity of the scattered beam is measured. It is
given by (eqs.~\ref{eq.rot_p1},~\ref{eq.th_int})
\begin{equation}
  I^\prime 
  \propto 
  S + P_1 \cos(2\eta) + P_2 \sin(2\eta),
  \label{eq.pa_eta}
\end{equation}
where
$S=({1+x^2})/({1-x^2})=({1+\cos^2(2\theta)})/({\sin^2(2\theta)})$ is
characteristic of the spillover; it is usually determined by measuring
a beam of known polarization, e.g., the incident beam or a strong
structural reflection of the sample.

%\subsubsection{Practical aspects}

In addition to the usual benefits of analyzer crystals, such as rejection of
diffuse scattering and fluorescence as well as further reduction of
higher harmonics of the X-ray wavelength, PA offers further advantages
which are specific to magnetic and resonant scattering.

Magnetic X-ray scattering at synchrotron sources is mostly done in the
vertical scattering geometry with incident $\sigma$ polarization. At
dipole resonances, all magnetic scattering is the $\pi$ polarization channel. 
A polarization analyzer set to accept only the $\pi$ channel thus
greatly improves the signal-to-noise ratio by preferentially
suppressing Thomson scattering, which remains $\sigma$
polarized. 

\begin{table*}

\caption[PA Crystals]{%
  A selection of crystals that can be used for polarization analysis. 
  The photon energy at which $2\theta_{\mathrm{PA}} = 90^\circ$ is 
  presented in column 3, while columns 4, 5 and 6 list some 
  absorption edges for which the corresponding crystal represents a good 
  choice.} 

\label{tab.pacrystals}
 
\centerline{%
\begin{tabular}{llr@{.}lr@{.}llll} 
\hline
\hline
Crystal & (H K L)    & \multicolumn{2}{c}{d}        
& \multicolumn{2}{c}{$E(90^\circ)$} & \multicolumn{3}{c}{Edges} \\  \
        &            & \multicolumn{2}{c}{[\AA]}        
& \multicolumn{2}{c}{[keV]} & 
$5f$ M$_{4,5}$ & $4f$ L$_{1,2,3}$ & $3d$ K\\  \hline 
Au       &  (1 1 1)   &  2&355 &  3&72 &  $ \mathrm{U,Np}$ & & \\ 
Pt       &  (1 1 1)   &  2&266 &  3&87 &                   & & \\
Cu       &  (2 0 0)   &  1&807 &  4&85 &  & & $\mathrm{Ti}$ \\
Graphite &  (0 0 4)   &  1&677 &  5&22 &  & & $\mathrm{Ti, V}$\\
Mo       &  (2 0 0)   &  1&574 &  5&57 &  & $\mathrm{La, Ce}$ & 
                                            $\mathrm{V}$ \\ 
Al       &  (2 2 0)   &  1&432 &  6&12 &  & $\mathrm{La, Ce, Pr, Nd}$ & 
                                            $\mathrm{Cr, Mn}$\\ 
Cu       &  (2 2 0)   &  1&276 &  6&86 &  & $\mathrm{Sm, Nd, Eu}$ & 
                                            $\mathrm{Mn, Fe}$\\ 
Au       &  (2 2 2)   &  1&177 &  7&44 &  & $\mathrm{Sm, Eu, Gd}$ &
                                            $\mathrm{Fe, Co}$\\ 
Al       &  (2 2 2)   &  1&169 &  7&49 &  & $\mathrm{Sm, Eu, Gd}$ &
                                            $\mathrm{Fe, Co}$\\
Pt       &  (2 2 0)   &  1&133 &  7&74 &  & $\mathrm{Sm, Eu, Gd, Tb, Dy}$ &
                                            $\mathrm{Co}$\\ 
Graphite &  (0 0 6)   &  1&118 &  7&84 &  & $\mathrm{Eu, Gd, Tb, Dy, Ho}$ &
                                            $\mathrm{Co}$\\ 
Cu       &  (2 2 2)   &  1&042 &  8&41 &  & $\mathrm{Gd, Tb, Dy, Ho, Er, Tm}$ &
                                            $\mathrm{Ni, Cu}$\\
Pt       &  (4 0 0)   &  0&981 &  8&94 &  & $\mathrm{Tb, Dy, Ho, Yb}$ &
                                            $\mathrm{Cu}$\\
Pd       &  (4 0 0)   &  0&973 &  9&01 &  & $\mathrm{Tb, Dy, Ho, Yb}$ &
                                            $\mathrm{Cu}$\\
Graphite &  (0 0 8)   &  0&839 & 10&48 &  & $\mathrm{Tm, Yb, Lu}$ &
                                            $\mathrm{Zn, Ga}$\\
Au       &  (3 3 3)   &  0&785 & 11&16 &  & & $\mathrm{Ge}$\\ 
\hline\hline
\end{tabular} 
}
\end{table*} 

For practical applications some additional issues have to be
considered. As was already pointed out, eq.~\ref{jones-thomson}
neglects multiple scattering effects. Diffraction from near-perfect
crystals where the dynamical theory of X-ray diffraction has to be
applied will give rise to deviations from this behavior. In extreme
cases, like the Renninger effect, the polarization dependence is
totally different \cite{Shen90,Shen92,Stetsko00}. Furthermore,
near-perfect crystals have very sharp rocking curves so that the
instrumental resolution function is strongly modified upon rotating the PA
about the scattered beam: In the $\sigma$ position, the acceptance
will be narrow in the $2\theta$ and wide in the $\chi$ directions of
a normal 4-circle diffractometer \cite{Busing67}, whereas in the $\pi$ position the
opposite is true. For some experiments it is desirable to improve the
longitudinal resolution. However, it is generally not convenient to
work with varying resolution functions. In particular, measurements of
the Poincar\'e-Stokes parameters, which require absolute values of the 
integrated intensity at several positions of the PA, have to be carried 
out with care. 

In principle, it is possible to determine $P_1$ ($P_2$) from a pair of
measurements at $\eta = 0^\circ$ and $90^\circ$ ($\eta=\pm 45^\circ$)
\cite{Vaillant77,Paixao02}. In order to estimate the
systematic errors, however, we recommend a fit of eq.~\ref{eq.pa_eta} 
to measurements at different positions of $\eta$ between $0^\circ$ 
and $180^\circ$ in steps of $30^\circ$ or less. 
Furthermore, the intensities should be recorded by rocking
the \emph{analyzer crystal} rather than the sample \cite{Detlefs04}; in this way
artifacts from the variable resolution function are minimized.

Finally, the sharp rocking curves require very high accuracy of the
alignment, mechanical precision and software control during changes of
the photon energy or the polarization channel.
Therefore crystals with a moderate mosaic width on the order of
$0.1^\circ$, for example metals, are preferred for most practical
applications. In this case the resolution function is defined by the
setting of slits before the PA assembly which do not rotate with the
PA\@. A serious disadvantage of such crystals is their relatively low
reflectivity, on the order of 1--10\%. The requirement of a fixed 
scattering angle $2\theta_{\mathrm{PA}}\approx 90^\circ$ leads to a 
linear increase of the momentum transfer, $Q$, with increasing photon energy,
and thus to diminishing form factors $f(Q)$. Thus, in general, the reflectivity 
of polarization analyzers decreases towards higher photon energies.
A list of some selected PA crystals commonly used, together with the X-ray absorption
edges they may be applied to, is presented in
table~\ref{tab.pacrystals}.

The selectivity of a polarization analyzer can be enhanced by reflecting 
the beam multiple times, e.g.~using a channel-cut crystal. The throughput, 
i.e.~the maximum transmitted intensity in the allowed polarization channel, 
however, will drop drastically unless perfect crystals such as \chem{Si} or 
\chem{Ge} are used.

Recently, a polarization purity of $2 \cdot 10^{-9}$ was achieved using channel-cut 
\chem{Si} crystals with 4 reflections \cite{Marx11} with a photon energy tuning to 
fulfill the $2\theta = 90^{\circ}$ condition as close as possible. The polarization purity in that 
experiment was limited by multiple-beam diffraction effects (Renninger effect) 
that partially rotate the plane of polarization and thus contaminate the 
``forbidden'' channel.

%%%%%%%%%%%%%%%%%%%%%%%%%%%%%%%%%%%%%%%%%%%%%%%%%%%%%%%%%%%%%%%%%%%%%%%%%%%%%%%%%%%

\section{Polarization analysis of resonant scattering at the \chem{Ge} K-edge}
\begin{figure}
	\begin{center}
	\begin{tabular}{cc}
		\parbox{0.540\columnwidth}{
			\includegraphics[width=0.540\columnwidth]{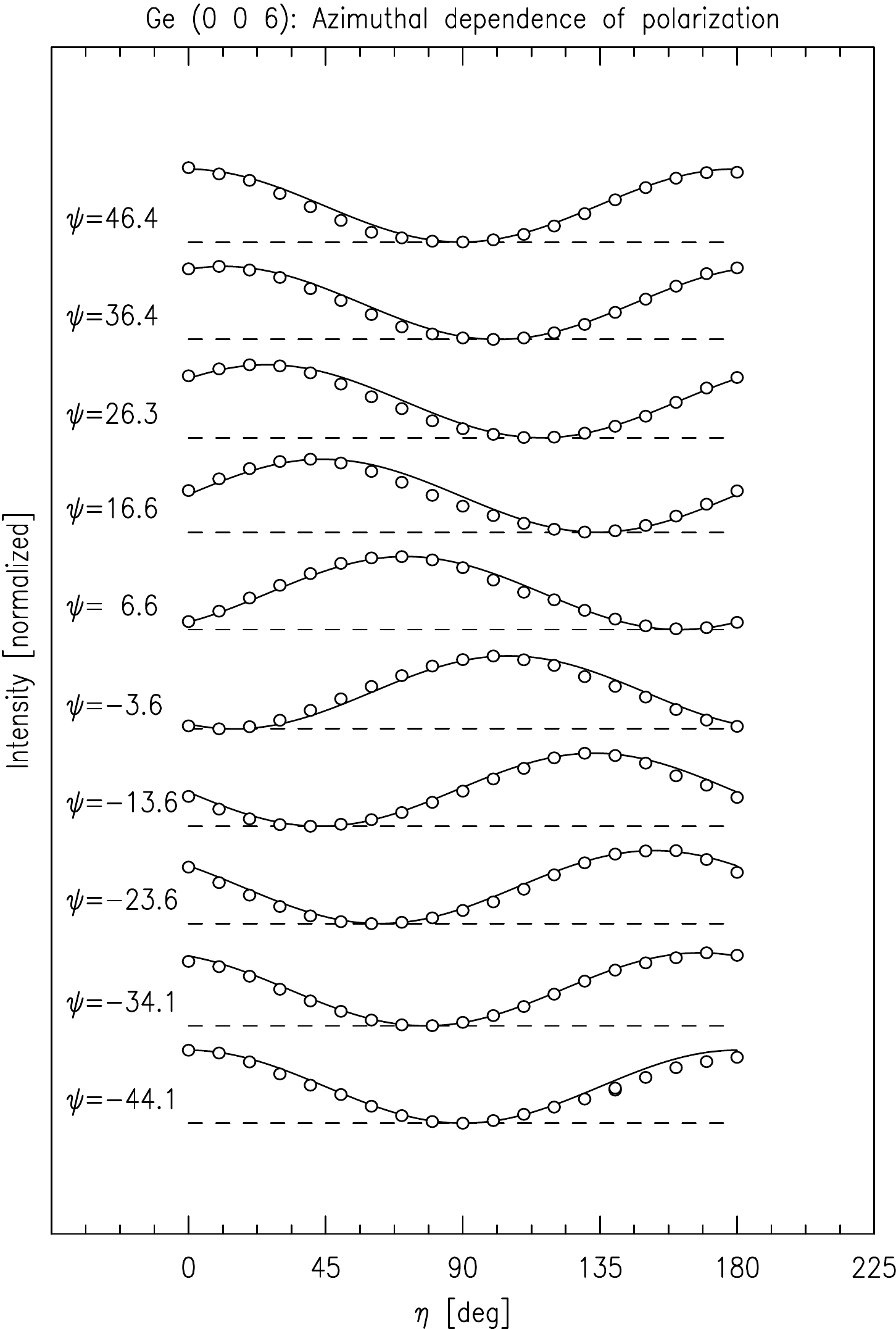}
		}
		&
		\parbox{0.390\columnwidth}{
			\includegraphics[width=0.390\columnwidth]{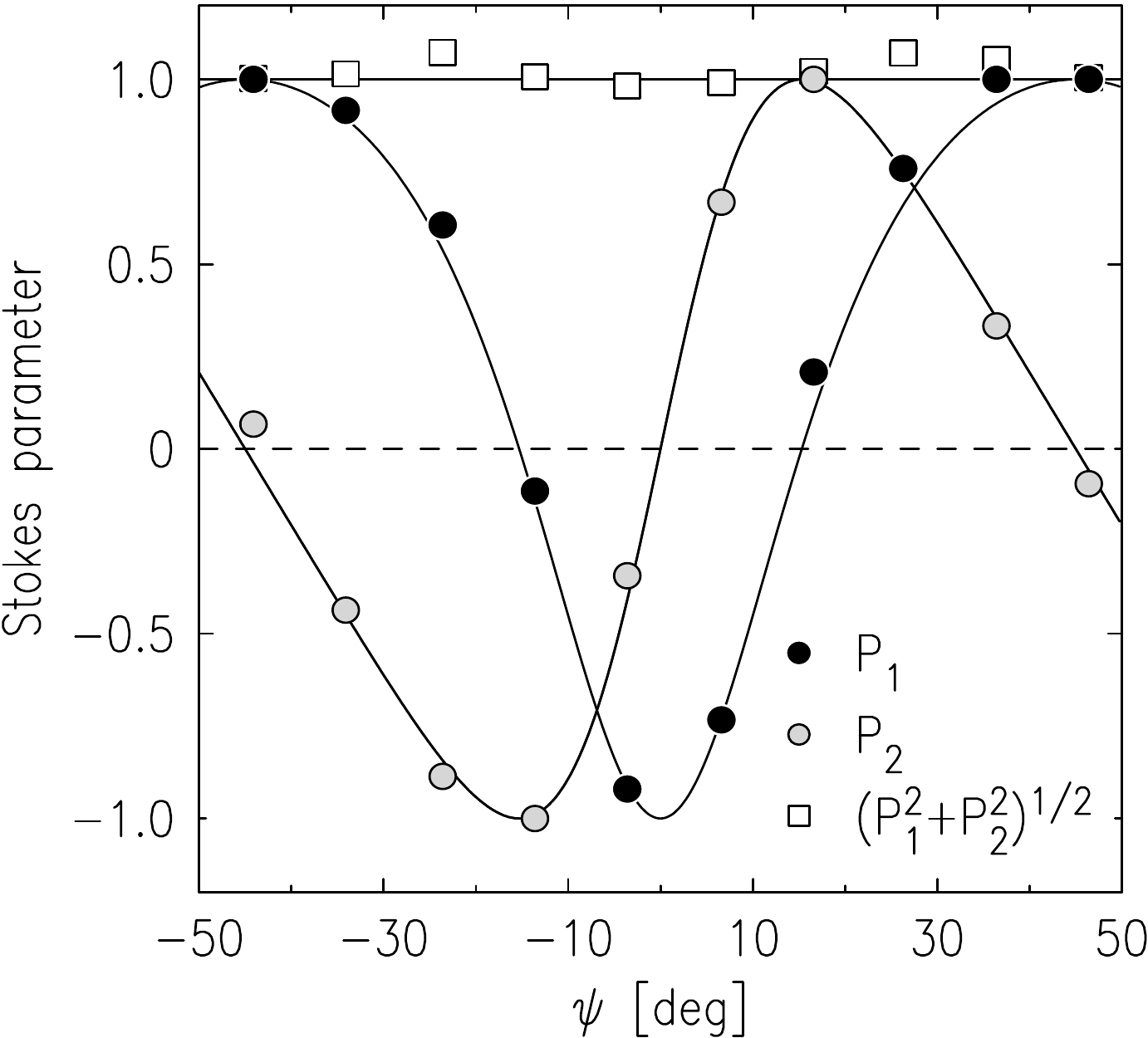}\\[5mm]
			
			\parbox{0.390\columnwidth}{
				\caption{Left: \label{fig_ge_cascade}Measurement of the Poincar\'e-Stokes 
					parameters as function of the azimuthal angle at the \chem{Ge} 
					$(006)$ reflection. Solid lines are fits to eq.~\ref{eq.pa_eta}. 
    			Right: \label{fig_ge_stokes}Poincar\'e-Stokes parameters extracted 
					from the measurements shown on the left by fitting 
					to eq.~\ref{eq.pa_eta}. The 
					experimental values are in quite satisfactory agreement with the 
					parameter-free predictions (eqs.~\ref{p1theo} and~\ref{p2theo}), 
					show as lines. Taken from~\cite{Detlefs04}.
				}
			}
		}
	\end{tabular}
	\end{center}
\end{figure}

%\begin{figure}
%	\centerline{%
%		\includegraphics[width=0.75\columnwidth]{fig_ge_cascade}
%	}
%	\caption{\label{fig_ge_cascade}Measurement of the Poincar\'e-Stokes 
%		parameters as function of the azimuthal angle at the \chem{Ge} 
%		$(006)$ reflection. Solid lines are fits to eq.~\ref{eq.pa_eta}. 
%		Taken from \cite{Detlefs04}.}
%\end{figure}
%
%\begin{figure}\sidecaption
%	\mbox{%
%		\includegraphics[width=0.45\columnwidth]{fig_ge_stokes}
%	}
%	\caption{\label{fig_ge_stokes}Poincar\'e-Stokes parameters extracted 
%		from the measurements shown in Fig.~\ref{fig_ge_cascade} by fitting 
%		to eq.~\ref{eq.pa_eta}. The 
%		experimental values are in quite satisfactory agreement with the 
%		parameter-free predictions (eqs.~\ref{p1theo} and~\ref{p2theo}), 
%		show as lines. Taken from~\cite{Detlefs04}.}
%\end{figure}

As an example of polarization analysis we present  the K-edge resonant scattering 
at the $(006)$ reflection of \chem{Ge} 
\cite{Templeton94,Kokubun01,Elfimov02,Elfimov02b,Detlefs04}. This 
reflection is crystallographically forbidden due to a glide plane 
extinction rule., i.e. the structure factor $F(\vec{Q})$ 
vanishes for scalar (Thomson) scattering. 
Rank-3 anisotropic tensor scattering (ATS), however, is allowed \cite{Templeton94}. 
Due to the high symmetry of \chem{Ge}, the ATS cross section is completely determined 
by symmetry, up to a (resonant) amplitude factor that varies as the photon energy is 
tuned across the resonance. At each given photon energy $\hbar \omega$ the polarization 
of the scattered beam as function of the the azimuth, $\psi$ (i.e. upon rotation of the 
crystal about the scattering vector $\vec{Q}$ \cite{Warren69}) is completely free of 
adjustable parameters. For incident $\sigma$ polarization the Poincar\'e-Stokes 
parameters of the scattered beam are given by \cite{Elfimov02,Elfimov02b,Detlefs04},
\begin{eqnarray}
  P_1(\theta,\psi)
  & = & 
  \frac{
    \sin^2(2\psi) - \sin^2(\theta)\cos^2(2\psi)
    }{
    \sin^2(2\psi) + \sin^2(\theta)\cos^2(2\psi)
    }
  \label{p1theo}
  \\
  P_2(\theta,\psi)
  & = &
  \frac{
    \sin(4\psi)\sin(\theta)    
    }{
    \sin^2(2\psi) + \sin^2(\theta)\cos^2(2\psi)
    }
  \label{p2theo}
  \\
  P_3(\theta,\psi)
  & = &
  0,
\end{eqnarray}
where $\theta$ is the Bragg angle.
In particular, we note that $P_1^2 + P_2^2=1$ for all azimuths, i.e.~the scattered 
beam is always fully linearly polarized. 

Experimental data were taken as follows \cite{Detlefs04}: 
The photon energy was tuned to the maximum of the resonance, $11.096\un{keV}$. 
For each setting of the azimuth, the \chem{Ge} $(006)$ reflection was aligned. 
Then the polarization analyzer was rotated about the beam in 
steps of $10^\circ$. At each position the integrated intensity was recorded 
by rocking the \chem{Au}(333) analyzer crystal. These integrated intensities 
were then fit to eq.~\ref{eq.pa_eta} in order to extract the 
Poincar\'e-Stokes parameters, see Fig.~\ref{fig_ge_cascade} (left). 
The final results are shown in Fig.~\ref{fig_ge_stokes} (right), 
along with the predictions of eqs.~\ref{p1theo} and~\ref{p2theo}. The agreement 
between the experimental data and the theoretical predictions is quite satisfactory.

Unfortunately, the polarization dependence is identical for the two possible 
microscopic mechanisms of the resonance, an E1--E2 mixed resonance 
\cite{Templeton94,Elfimov02,Elfimov02b}, or a E1 resonance 
of atoms displaced from their equilibrium positions by thermal motion 
(thermal motion induced scattering, TMI) \cite{Kokubun01}. 
Therefore this experiment was unable to distinguish between the two competing models.
Studies of the amplitude and line shape as function of temperature, however, favor 
the TMI model \cite{Kokubun01,Oreshko11}. 

%%%%%%%%%%%%%%%%%%%%%%%%%%%%%%%%%%%%%%%%%%%%%%%%%%%%%%%%%%%%%%%%%%%%%%%%%%%

\section{Complete characterization of the polarization state}

The complete determination of the polarization state of a synchrotron beam is
essential for characterizing the circular polarized light produced by using 
X-ray phase plates or exotic insertion devices. Two main methods have been 
successfully used for this purpose.

The first is based on the same principle as used in the visible region: A full 
characterization of the polarization of a wave (in terms of the Poincar\'e-Stokes vector) 
can be obtained by using the combination of a phase retarder and a linear 
polarization analyzer \cite{Born99}. In the X-ray regime, this was first 
proposed by \cite{Ishikawa91} and then used by \cite{Giles95}.

The second polarimetry method is to use multiple beam diffraction. The phase 
probing capability of this method can allow for the unambiguous determination 
of the Poincar\'e-Stokes parameters and can be applied to X-ray 
polarimetry \cite{Shen92,Shen93}.

%%%%%%%%%%%%%%%%%%%%%%%%%%%%%%%%%%%%%%%%%%%%%%%%%%%%%%%%%%%%%%%%%%%%%%%%%%%

\begin{acknowledgement}
The authors thank F. de Bergevin, T. Roth, 
and S.~B.~Wilkins for many stimulating discussions.
The ESRF is acknowledged for provision of beam time on beamline ID20.
\end{acknowledgement}

\bibliographystyle{epj}
%\bibliography{../../allref,specialref}
\bibliography{polarization_I}
\end{document}